\documentclass[%
 reprint,
superscriptaddress,
longbibliography,
 amsmath,amssymb,
 aps,
 prb,
]{revtex4-2}

\usepackage[T1]{fontenc} 
\usepackage{graphicx}
\usepackage{amsmath}
\usepackage{amssymb}
\usepackage{textgreek}
\usepackage{tabularx}

\usepackage[usenames]{color}
\definecolor{redcolor}{RGB}{255,0,0}
\definecolor{bluecolor}{RGB}{0,100,255}
\definecolor{blackcolor}{RGB}{0,0,0}

\newcommand{\dCo}{d_\mathrm{Co}}
\newcommand{\mx}{\langle m_x\rangle}
\newcommand{\mz}{\langle m_z\rangle}
\newcommand{\Aex}{A_\mathrm{ex}}
\newcommand{\Hspar}{H_\mathrm{s}^{||}}
\newcommand{\Hpar}{H^{||}}
\newcommand{\Hsperp}{H_\mathrm{s}^{\perp}}
\newcommand{\Ms}{M_\mathrm{S}}
\newcommand{\kc}{k^*}
\newcommand{\pc}{p^*}

\begin{document}

\title{Spin wave freezing in Re/Co/Pt multilayers}

\author{Jan Kisielewski}\email{jankis@uwb.edu.pl}
\affiliation{Faculty of Physics, University of Bia\l ystok, Bia\l ystok, Poland}
\author{Kilian Lenz}
\affiliation{Institute of Ion Beam Physics and Materials Research, Helmholtz-Zentrum Dresden-Rossendorf, Germany}
\author{Pawel Gruszecki}
\affiliation{Faculty of Physics, Adam Mickiewicz University, Pozna\'{n}, Poland}
\author{Ryszard Gieniusz}
\affiliation{Faculty of Physics, University of Bia\l ystok, Bia\l ystok, Poland}
\author{Urszula Guzowska}
\affiliation{Faculty of Physics, University of Bia\l ystok, Bia\l ystok, Poland}
\author{Marek Kisielewski}
\affiliation{Faculty of Physics, University of Bia\l ystok, Bia\l ystok, Poland}
\author{Artem Lynnyk}
\affiliation{Institute of Physics Polish Academy of Sciences, Warsaw, Poland}
\author{Aleksiej Pietruczik}
\affiliation{Institute of Physics Polish Academy of Sciences, Warsaw, Poland}
\author{Andrzej Wawro}
\affiliation{Institute of Physics Polish Academy of Sciences, Warsaw, Poland}
\author{Andrzej Maziewski}
\affiliation{Faculty of Physics, University of Bia\l ystok, Bia\l ystok, Poland}

\begin{abstract}

The phenomenon of spin wave (SW) freezing occurs in the Damon-Eshbach mode in thin film magnetic systems, when SW phase and group velocities both go to zero, and the wave ceases to oscillate and move, preserving its shape as a domain structure pattern. This effect is related to the spin reorientation transition, where the magnetization configuration changes between the homogeneous in-plane state and domain structure with the out-of-plane magnetization component state. Here, we study the SW freezing effect in [Re/Co/Pt]$_{20}$ magnetic multilayers, induced by varying the in-plane external magnetic field. The studies were performed on nanostructures with the quality factor $Q$ (ratio of uniaxial anisotropy to demagnetization energies) greater and smaller than one. Domain structures with an out-of-plane magnetization component were observed in these multilayers. The critical field, visible as the saturation field $\Hspar$ in the parallel static magnetization curve measured by superconducting quantum interference device (SQUID), is also manifested in the field-dependent vector-network-analyzer ferromagnetic resonance (VNA-FMR) experiment, which measures the homogeneous magnetization oscillations. Brillouin Light Scattering (BLS) spectra, recorded for several values of wave vectors and several field values, probed the field-evolution of the dispersion relation. Micromagnetic simulations allow one to obtain a full dispersion, in good agreement with VNA-FMR and BLS results. Around $\Hspar$ the simulated dispersion relations approach the conditions for SW freezing. Below $\Hspar$ low and high frequency VNA-FMR modes are related to magnetization oscillations inside domain walls and within domains, respectively. The experimental results of static and dynamic behavior, together with micromagnetic simulations, create an overall consistent picture of the investigated multilayers. 

\end{abstract}

\maketitle

\section{Introduction}

Spin wave softening---the reduction of magnon frequency toward zero at a finite wave vector---is a fundamental indicator of magnetic instabilities and phase transitions. When phase and group velocities both vanish simultaneously, spin wave (SW) ceases to propagate and oscillate and a phenomenon termed SW freezing occurs \cite{Leaf06, Ki23, Gallardo25, Cep26, Battistelli2025arxiv}, which marks the critical point below which an instability develops \cite{Sobucki25} and drives the nucleation of a static stripe-domain structure. The connection between soft modes and domain nucleation has been demonstrated across diverse platforms: in ferromagnetic nanostructures \cite{Leaf06}, thin films with perpendicular magnetic anisotropy (PMA) \cite{Ki23}, where the softening produces characteristic sombrero- or cowboy-hat-shaped dispersion relations of SWs \cite{lesniewski26}, and synthetic antiferromagnetic multilayers where the instability can become direction-selective \cite{Gallardo25}. Moreover, in systems with nonreciprocity induced by Dzyaloshinskii-Moriya interaction (DMI), a slow-instability regime exists between the exceptional-point field  and the critical field for stripe domain formation, where Gilbert damping paradoxically enhances SW amplitude rather than suppressing it \cite{Sobucki25}. On the other hand, stabilized stripe domains offer a unique platform for studying SW dynamics, as---unlike lithographically defined magnonic crystals---they can be continuously tuned by external magnetic field, enabling reversible control of SW propagation without nanofabrication \cite{Sz22, Ba17, Li15, Gr22,Dhiman24}. Such stripe-based magnonic crystals can exhibit nonreciprocal dispersion relations, originating either from dipolar coupling \cite{Sz22} or from interfacial DMI \cite{Dhiman24}. 

In our recent paper \cite{Ki23} we described theoretically the effect of SW freezing. Experimental realization of this phenomenon poses several challenges. Magnetic parameters such as anisotropy and DMI are typically coupled through their common dependence on film thickness and interface quality \cite{Al92, Ki02, Maziewski2014, Belabbes2016, Dhiman21}. Multilayer systems, where ultrathin magnetic layers are separated by nonmagnetic spacers, offer a route to address this problem: the repeated interfaces enhance both PMA and interfacial DMI, while the total magnetic thickness can be tuned independently to control demagnetizing effects \cite{Wo16, Ja20}. In such systems, short-range interface-originated interactions coexist with long-range dipolar coupling, resulting in weak stripe domains similar to those in thick films---out-of-plane magnetic domains separated by hybrid-type domain walls with flux-closure caps \cite{Ki23, Hubert98}. Depending on the layer thickness and repetition number, multilayers can host different domain configurations---from tilted stripe patterns in thinner systems to well-defined up/down domains in thicker ones.

Whereas micromagnetic simulations provide complete access to magnetization dynamics---amplitude $A$ and phase $\Phi$ as functions of wave vector $k$, frequency $f$, magnetic field $H$, and spatial position $\vec x$, i.e., the full dependence $A(k, f, H, \vec x)$ is readily accessible---experimental techniques probe only limited subspaces of this information. The most established method is ferromagnetic resonance (FMR), where the sample is excited by an ac magnetic field of microwave frequency (typically $f_0$ = 9.5 GHz in x-band spectrometers) and the resonance is detected by sweeping the dc magnetic field, thus probing only $A(k = 0, f = f_0, H)$. Vector-network-analyzer FMR (VNA-FMR) extends this approach by sweeping both $f$ and $H$, yet still measures only the uniform mode ($k = 0$). Brillouin light scattering (BLS), employing inelastic magnon scattering, accesses finite wave vectors by adjusting the incident angle $\theta$ of the laser beam according to $k = 4\pi \sin \theta/\lambda$. Sweeping $\theta$ and $H$ yields the dispersion relation $A(k, f, H)$, with the additional advantage of spatial resolution down to the single-micrometer scale, providing $A(k, f, H, [x, y])$. Nevertheless, BLS has some limitations: low frequencies near the elastic peak and high wave vectors beyond the accessible angular range remain out of reach. The combination of these complementary techniques with simulations is therefore essential for a complete characterization of SW dynamics in stripe domain systems. FMR studies have revealed magnetization dynamics and oscillation modes in films with weak stripe domains \cite{Vu00, VuE00, Ta10, Ca17}, while BLS measurements have probed SW propagation and dispersion in stripe-based magnonic crystals \cite{Sz22, Ba17, Gi24} as well as localized modes of domains and domain walls \cite{Gu12, Ta14, Ca17}. 

Recently, we investigated SW propagation in Ir/Co/Pt multilayers with interfacial DMI, providing the first experimental study of SW dynamics in stripe domain patterns in films with DMI \cite{Gi24}, where the relatively low PMA and thin multilayer stack in that system resulted in hybrid stripe domains at remanence, characterized by a nonzero in-plane magnetization throughout the whole sample. The complementary regime of well-defined up/down domains, where magnetization points fully out of plane within each stripe, remained experimentally unexplored in terms of SW dynamics. To access that regime and investigate SW behavior near the critical field for SW freezing, we deliberately designed [Re/Co/Pt]$_{20}$ multilayers with thicker Co layers and a higher repetition number. This choice was motivated by the need to simultaneously enhance PMA and strengthen DMI through repeated asymmetric interfaces, while increasing the total magnetic thickness to stabilize the out-of-plane domain configuration. Such well-defined up/down domain structures at remanence can appear in a shape of stripe domains or a bubble lattice, while demagnetizing the system by the magnetic field applied at specific angles \cite{jena2026ass}. Resonant dynamics dependent on magnetic field history has been also discussed in these multilayers \cite{jena2026prb}.
By tuning the Co thickness, one can access systems with quality factor $Q$ [$Q=2K_\mathrm{u}/(\mu_0\Ms^2)$, where $K_\mathrm{u}$ is uniaxial magnetic anisotropy, and $\Ms$ the saturation magnetization] both above and below unity. 
Consequently, we can probe SW dynamics across the spin reorientation transition and, for the first time in multilayers with DMI, systematically characterize the SW freezing process near the critical field. 
We demonstrate field-induced SW freezing by combining complementary approaches. Static magnetization curves obtained by SQUID determine the saturation fields and magnetic anisotropies, while MFM confirms the stripe domain structure at remanence. VNA-FMR measurements trace the evolution of the uniform precession mode as the in-plane field approaches the critical value, and BLS spectroscopy probes the dispersion relation at finite wave vectors. Micromagnetic simulations reveal the full dispersion relation and visualize the freezing process. The consistent picture emerging from experiment and simulation provides experimental evidence for SW freezing near the spin reorientation transition.

\section{Methods}

The measurements of static magnetization curves were done by Quantum Design Magnetic Property Measurement System (MPMS XL7), equipped with an integrated superconducting quantum interference device (SQUID) magnetometer. The isothermal field dependences of magnetic moment were recorded at room temperature, by utilizing the Reciprocating Sample Option (RSO). The RSO provided oscillating movement for the studied samples through the SQUID pick-up coils, with a precision of about 10$^{-8}$ emu.

The magnetization dynamics were recorded by the vector-network-analyser ferromagnetic resonance (VNA-FMR) technique. The magnetic field was applied in the sample plane and the system was excited by an ac magnetic field at microwave frequencies. A coplanar waveguide (CPW) with an 80-µm-wide center conductor, aligned along the $y$ axis, was used for excitation; thus the microwave field had components perpendicular to the CPW axis, i.e.~the in-plane ($x$) and out-of-plane ($z$) components. The experimental configuration is presented in the inset in Fig.~\ref{figtwo}(b). 
At each frequency, the field was swept from -2~T to 2~T, and the frequency changed from 0 to 35~GHz. For each pair of $f$ and $\Hpar$ the response of the system was recorded as absorption of the initial microwave signal---strong absorption corresponds to the ferromagnetic resonance. The signal was averaged over the excited area of the sample, i.e.~above the signal line of the coplanar waveguide. So the result corresponds to the oscillations with $k=0$.

Brillouin light scattering (BLS) operated in backscattering configuration, enabling measurements of the SW dispersion relation for wave vectors $k$ between $\sim$4 and $\sim$23 rad/$\mu$m, corresponding to the incident angle $\theta$ of the laser beam, varied between 10$^\circ$ and 70$^\circ$ from the sample normal ($k = 4\pi \sin\theta/\lambda$, $\lambda = 532$~nm). The magnetic field was applied in the sample plane. The same objective was used to illuminate the sample and to collect the backscattered light, which was analysed by a TFPII Sandercock multi-pass tandem Fabry-Perot interferometer. 

The micromagnetic simulations were performed using \textsc{MuMax3} software \cite{Va14}. The geometry in the simulations was as follows. The system size was $N_x\times 1\times20$. One layer of cells corresponds to each of the 20 films of the multilayer system. The effective medium model was applied \cite{Wo16}---each magnetic layer together with the surrounding non-magnetic spacer film was treated as a single layer, with ``diluted'' magnetic properties, in proportion to the fraction of magnetic material within each layer. It should be emphasized, that all values of magnetic parameters quoted in the paper concern the pure, nondiluted magnetic layers, and proper scaling was done at the level of simulation scripts. The cell size was 1~nm $\times$ 1~nm $\times$ $C_z$, where $C_z$ was (1~nm + $\dCo$ + 1~nm). Along the $y$ direction there was only 1 cell, but using the periodic boundary conditions the system pretended an infinite plane. However any non-trivial magnetization distribution was allowed only in the $xz$ plane that corresponds to the stripe domain structure, with stripes along the $y$ direction. All dc magnetic fields were applied only along the $y$ direction.
In order to reasonably account for the effect of nonmagnetic spacers, which make the magnetic films barely interact with each other via exchange interaction, the interlayer exchange coupling was decreased to 10\% of the intra-layer value in our geometry with effective medium.

\section{Experimental results}

\begin{figure*}[tbh!]  
\centering
 \includegraphics[width=0.98\textwidth]{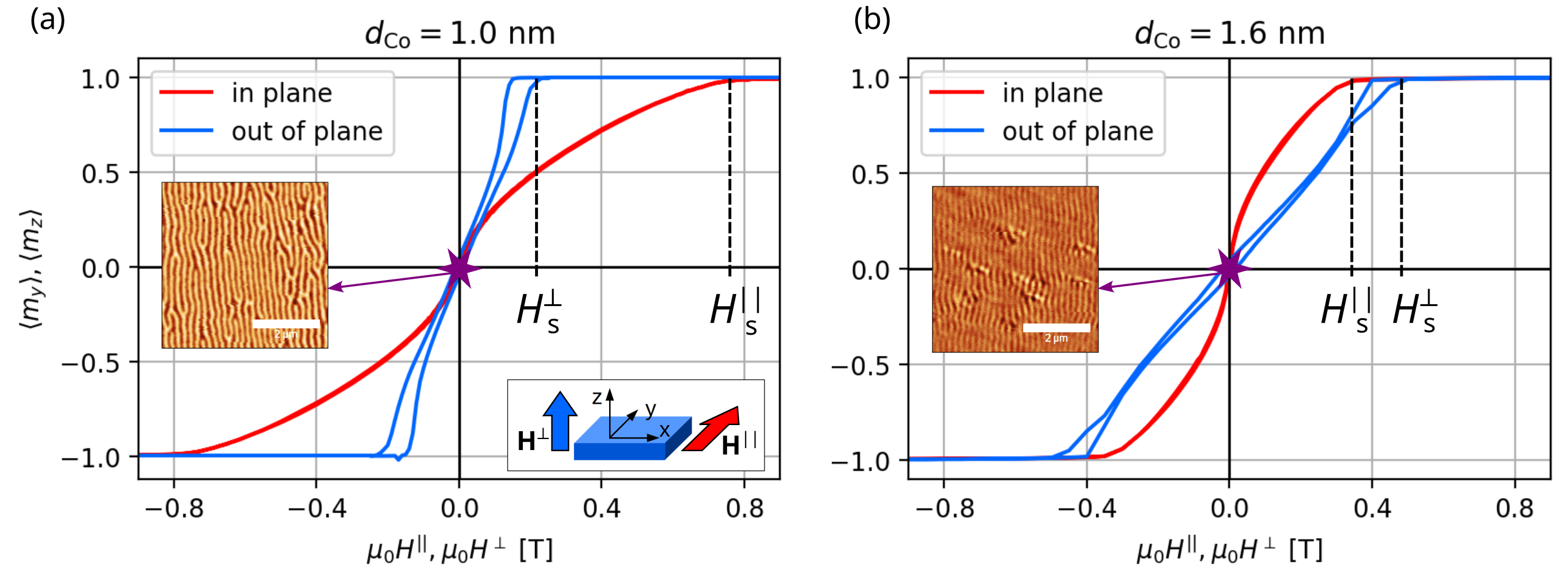}
\caption{Static magnetization curves measured by SQUID, for (a) $\dCo$ = 1.0~nm, (b) $\dCo$ = 1.6~nm, in in-plane (red curves) and out-of-plane (blue curves) configuration. Insets: MFM images of the domain structure at zero field.}
\label{figone}
\end{figure*}

The studied samples were produced with the molecular beam epitaxy (MBE) technique. Additionally the film quality was monitored in-situ with reflection high-energy electron diffraction (RHEED), and the structural quality was also confirmed by x-ray diffraction (XRD) and local transmission. The samples have a multilayered structure [Re (1~nm) /Co ($\dCo$) /Pt(1~nm)]$_{20}$. Some results on samples from the same series, including structural characterization and bubble lattice formation, was published recently~\cite{jena2026ass}. Enhanced spin pumping in asymmetric Pt/Co/Re/Co/Pt has also been reported \cite{bonda2026prb}. Two values of cobalt thicknesses $\dCo$ = 1.0 and 1.6~nm were used. 
Different thicknesses of the magnetic film are expected to yield different values of magnetic anisotropy \cite{Ki02,Dhiman25}. On the other hand, an asymmetric interface, with either Re or Pt neighborhood, results in significant DMI \cite{Ja20,Dhiman25}. Basic characterization of the samples was made by superconducting quantum interference device (SQUID) magnetometry, with in-plane and out-of-plane field configuration, at room temperature. The measured curves, presented in Fig.~\ref{figone}, exhibit different values of the saturation fields $H_\mathrm{s}$, namely $\Hspar$ and $\Hsperp$, for in-plane or out-of-plane field configurations. The relation between these fields reflects the magnetic anisotropy: in the case of $\dCo$ = 1.0~nm $\Hspar$ > $\Hsperp$, which suggests $Q>1$ (easy magnetization axis perpendicular to the film, lower field $\Hsperp$ saturates the magnetization along the easy out-of-plane direction), and the opposite dependence $\Hspar < \Hsperp$, and thus $Q<1$ (easy magnetization plane, film plane) for $\dCo$ = 1.6~nm. The values of anisotropy can be determined from the difference between the areas under the out-of-plane and in-plane magnetization curves \cite{Jo96}. In the case of out-of-plane curves, where some hysteresis was observed, the average of the two branches of hysteresis was taken into account. Thus, $Q=1.18$ and $Q=0.88$ for $\dCo=1.0$ and 1.6~nm, respectively. From the point of view of SW freezing, the values of $\Hspar$ are important, $\mu_0\Hspar$ = 0.76~T for $\dCo=1.0$~nm and 0.34~T for $\dCo=1.6$~nm. The SQUID measurement also allowed for determination of $\Ms$, which is 1.45 MA/m for both studied samples. 

The samples were also studied by magnetic force microscope (MFM) at zero field, after controlled magnetic history---before the measurement the samples had been saturated with sufficiently strong magnetic field applied in the sample plane along $y$ axis. MFM images, presented as insets in Fig.~\ref{figone}, reveal a stripe domain structure. The periods at remnant state were determined to be $p=170$ and 178 nm for $\dCo=1.0$ and 1.6~nm, respectively. 

\begin{figure*}[tbh!]  
\centering
 \includegraphics[width=0.98\textwidth]{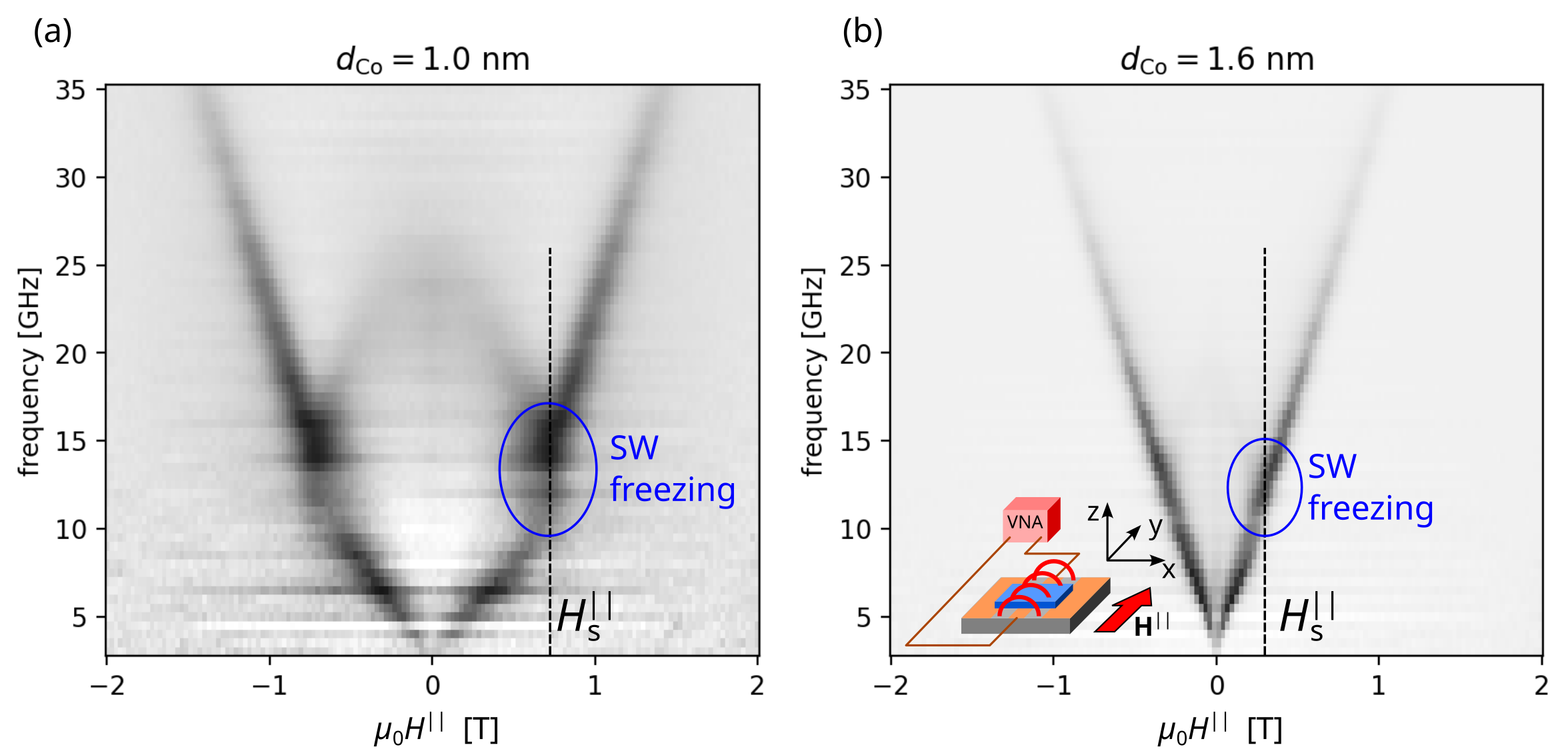}
\caption{Magnetization dynamics recorded by VNA for (a) $\dCo$ = 1.0~nm, (b) $\dCo$ = 1.6~nm. In-plane field and frequency dependence of the ferromagnetic resonance is visualized as gray-level. SW freezing is expected at $\Hspar$. Above and below this field the character of dynamics is different: a single band (Kittel mode) is observed for saturated homegenous magnetization, two bands with either low or high frequency are observed in the magnetic domain state. The inset in (b) presents a configuration of the VNA-FMR experiment.}
\label{figtwo}
\end{figure*}

\begin{figure*}[tbh!]  
\centering
 \includegraphics[width=0.98\textwidth]{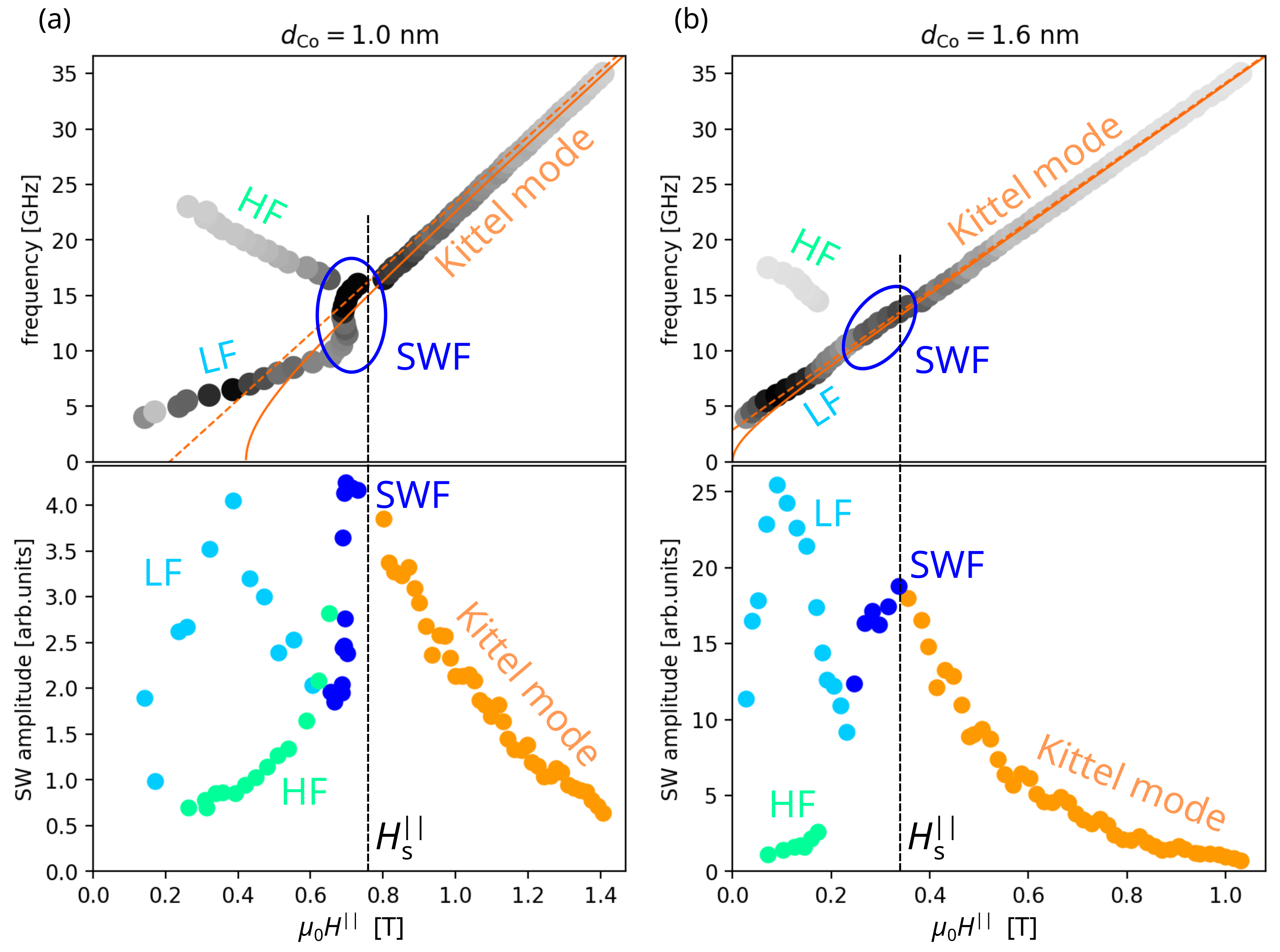}
\caption{Analysis of the resonance frequency (upper panels) and amplitude for (a) $\dCo=1.0$~nm, (b) $\dCo=1.6$~nm. Orange, green, light blue, and dark blue colors distinguish the particular branches: Kittel mode, high-frequency (HF), low-frequency (LF), and neighborhood of spin wave freezing (SWF), respectively. The orange continuous lines are the fits to the $f(\Hpar)$ dependences within the Kittel mode, the dashed lines are their asymptotes, which cross the field axis at either positive or negative values, according to the effective anisotropy field. }
\label{figthree}
\end{figure*}

The magnetization dynamics were recorded by the VNA-FMR technique. 
The resonant absorption spectra, related to the SW amplitude, are presented in Fig.~\ref{figtwo} as gray-scale maps $A(f,\Hpar)$. For both samples clear dark bands are visible, representing the resonances. The branches were analyzed in detail, by fitting the Lorentzian peaks to each field-swept spectra (horizontal scans across the maps in Fig.~\ref{figtwo}). The localization of each peak (i.e.~the resonant field for each studied frequency), as well as its amplitude, relative to the background, were obtained. The extracted data is presented in Fig.~\ref{figthree}. In the top panels there are plots in coordinates of in-plane field and frequency, with the peak amplitude marked with gray level (i.e. similarly as in Fig.~\ref{figtwo}). In the bottom panels the plot coordinates are field and SW amplitude. One can distinguish characteristic branches, marked with colors. For high magnetic field amplitudes above the $\Hspar$ values (derived also from the static magnetic measurement), the field-dependencies of the resonant frequency are monotonic (marked in orange). These are expected to correspond to the Kittel modes---the uniform oscillations of a homogeneously magnetized sample. The Kittel formula $f=\frac{\gamma \mu_0}{2\pi}\sqrt{\Hpar(\Hpar-H_\mathrm{ani})}$ (where $\gamma$ is the gyromagnetic ratio, $\Hpar$ is the external in-plane field, and $H_\mathrm{ani}=\Ms(Q-1)$ is the effective anisotropy field) can be fitted within this branch, using the values of $\Ms$ and $Q$ from the static experiment (see above), yielding an estimation of $\gamma$, $\gamma=175$ and 192~GHz/T for $\dCo=1.0$ and 1.6~nm, respectively. Slightly different values for different Co layer thicknesses are not surprising \cite{Sc56}. The fitted lines are plotted in the top panels of Fig.~\ref{figthree} with orange solid lines. Their high-field asymptotes (plotted with dashed orange lines) cross the horizontal axes at either positive (for $\dCo=1.0$~nm) or negative ($\dCo=1.6$~nm) values, according to the sign of the effective anisotropy field values. 

In this field regime, it is also important to note the field-dependent amplitude, which increases strongly while reducing the field down to $\Hspar$. This overall increase in amplitude for broad range of wave vectors (including $k=0$, the case of VNA-FMR measurements), is expected for the SW freezing effect \cite{Ki23}. 
Below $\Hspar$, two branches are visible, with low (marked in blue) or high frequencies (marked in green). In this field range the sample is in the domain state, so the particular branches are expected to correspond to oscillations within components of the domain structure. Interestingly, the intensity of the low-frequency branch is much higher, moreover its amplitude exhibits a nonmonotonic field-dependence, with a distinct maximum at moderate fields. The transient branch around $\Hspar$ (marked in dark blue), connected with SW freezing, has a high amplitude.

\begin{figure*}[tbh!]  
\centering
\includegraphics[width=0.98\textwidth]{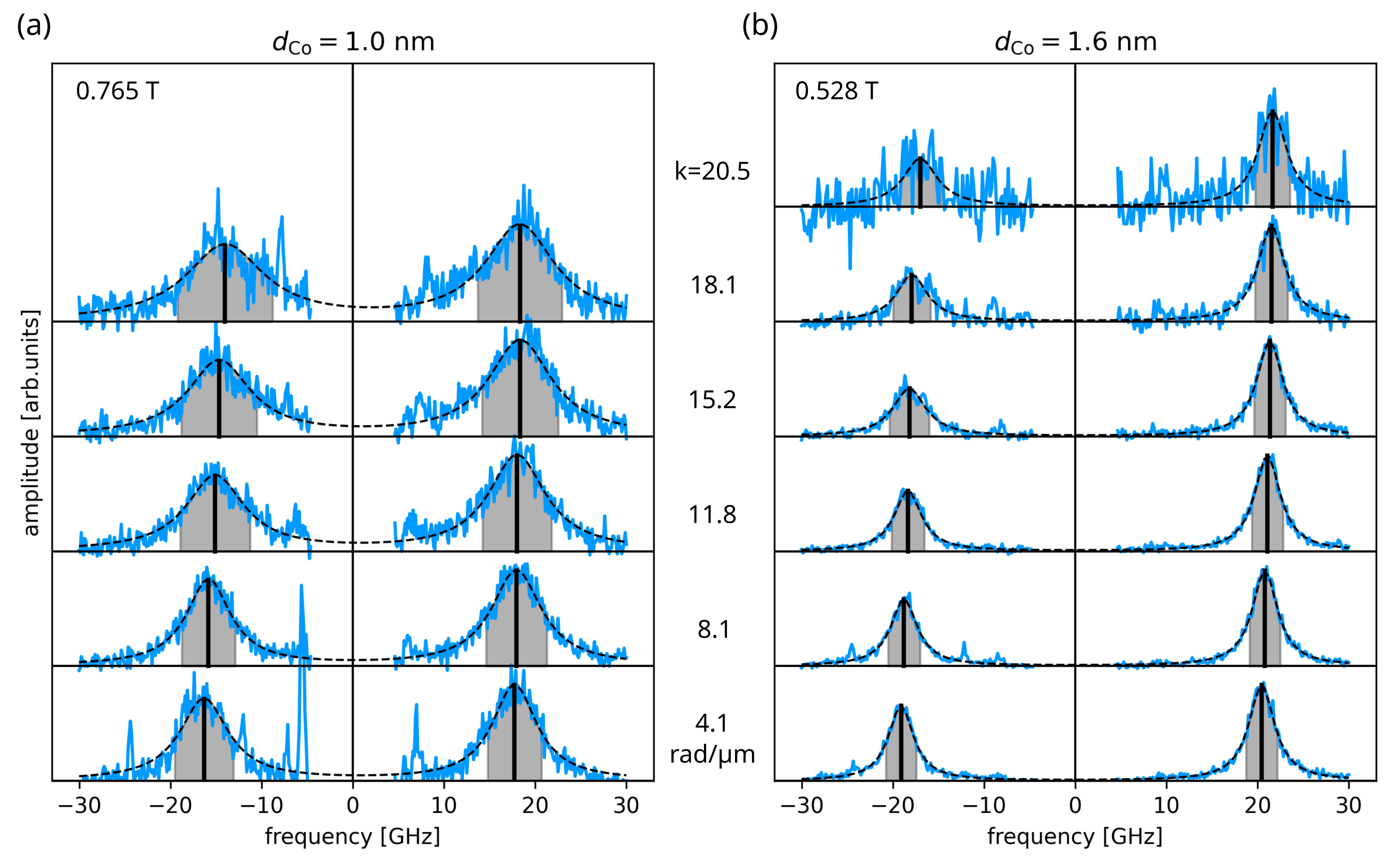}
\caption{BLS spectra for (a) $\dCo=1.0$~nm, recorded at $\mu_0\Hpar$ = 0.765~T, (b) $\dCo=1.6$~nm, recorded at $\mu_0\Hpar$ = 0.528~T. In both cases the field values are above $\Hspar$, so the samples are homogeneously magnetized. The spectra plotted in blue were recorded for different values of $k$, as labeled in the space between the subplots. The peaks were detected by fitting the Lorentzian function (black dashed lines). The central peak frequency values are marked by black vertical lines, the FWHM values are visualized by gray shading.}
\label{figfour}
\end{figure*}

BLS spectra were measured at in-plane magnetic fields, for a series of $k$ values in the range from 4.1 to 20.5 rad/$\mu$m. To obtain the frequencies of Stokes $f_S$ and anti-Stokes $f_{aS}$ peaks, the Lorentzian function was fitted to each BLS spectrum. The series of $k$-dependent spectra are presented in Fig.~\ref{figfour}. The determined frequencies and peak full-width at half-maximum (FWHM) values are marked as thick vertical black lines, with gray shading around them. The clearly visible asymmetry of the Stokes and anti-Stokes points is consistent with the presence of DMI discussed in single Co layers with Pt and Re surrounding \cite{Dhiman25,Fa24}.
The field-dependent evolution of the dispersion relation, its physical interpretation, and the quantitative comparison with micromagnetic simulations, which provide access to wave vectors beyond the BLS range and allow connection to the SW freezing mechanism, are discussed in detail in the following Section.

\section{Discussion}

To further interpret these observations, establish some links between them, and draw valuable conclusions, we have performed micromagnetic simulations of the magnetization statics and dynamics, using the \textsc{MuMax3} software \cite{Va14}, assuming the bias magnetic field is applied along the $y$ axis. The details of configuration and geometry of the simulated system are described in the Methods section. In the simulations the experimental values of $\Ms$ and $Q$ were implemented. The DMI values, $D=3.23$ and 2.18~mJ/m$^2$ for $\dCo=1.0$ and 1.6~nm, respectively, were taken from DFT calculations, reported for the Re/Co/Pt systems elsewhere \cite{Fa24, Dhiman25}. In preliminary simulations, the magnetization was relaxed in zero field for several values of the exchange constant $\Aex$. For each, the total energy density of the system was minimized in order to determine the energetically preferred period. The $\Aex$-dependence of the minimum-energy period was validated using the experimental periods, measured by MFM. In this way, the best-fit value of $\Aex$ = 23~pJ/m was found for both samples, while the out-of-plane exchange was reduced to $0.1\cdot\Aex$ as often done to model multilayers in micromagnetic simulations \cite{Gi24}.

\begin{figure*}[tbh!]  
\centering
\includegraphics[width=0.98\textwidth]{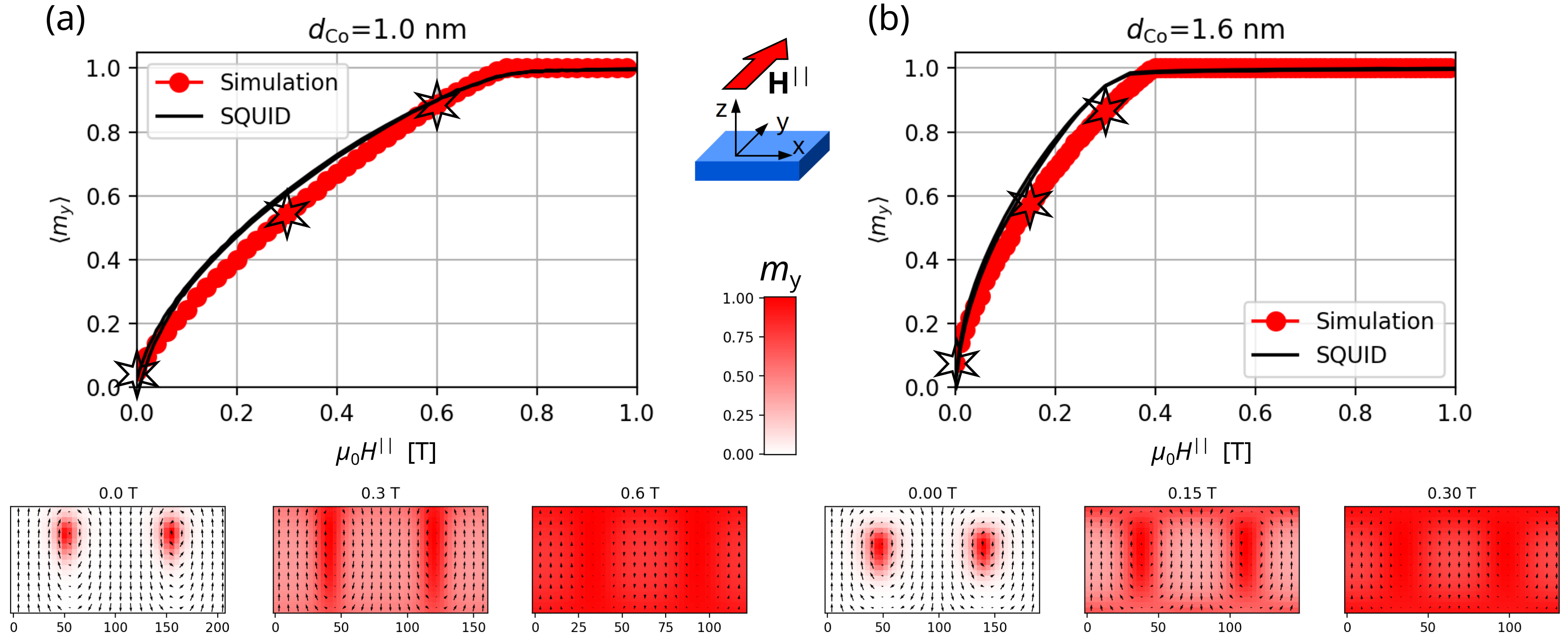}
\caption{Simulated static magnetization curves, for (a) $\dCo=1.0$~nm, (b) $\dCo=1.6$~nm. For comparison, the results of the SQUID measurements are plotted (black lines). For each sample, the calculated magnetization distribution at three selected field values (marked with asterisks) is presented at the bottom. The arrows denote the local magnetization orientation in the $xz$ plane, the red color intensity represents the $m_y$ component. }
\label{figfive}
\end{figure*}

Having established the magnetic parameters, static magnetization curves were calculated by relaxing the system in gradually increasing in-plane magnetic field. The calculated curves are compared to the SQUID measurements in Fig.~\ref{figfive}. For both samples relatively good agreement is observed, with some discrepancy at moderate fields, where the experimental curves rise more gradually than the simulated ones. However, the saturation fields $\Hspar$ are well reproduced: $\mu_0\Hspar \approx$ 0.74 and 0.38~T for $\dCo = 1.0$~nm ($Q > 1$) and 1.6~nm ($Q<1$), respectively. The simulations reveal hybrid domain walls characteristic of systems with DMI: N\'eel caps appear at the top and bottom surfaces, with a Bloch-type core in the central region of the multilayer, slightly displaced from the mid-plane (i.e.,~not centered exactly at half-thickness) due to DMI \cite{Le18, Mo17}. For the thinner Co layer ($\dCo = 1.0$~nm), where the DMI is larger, this shift is more pronounced. The simulations also show that, for both samples, while increasing $\Hpar$ the equilibrium domain period decreases from 170--180~nm at zero field to about 120~nm -- the critical period $\pc$ while approaching the critical field $\Hspar$.

\begin{figure*}[tbh!]  
\centering
 \includegraphics[width=0.98\textwidth]{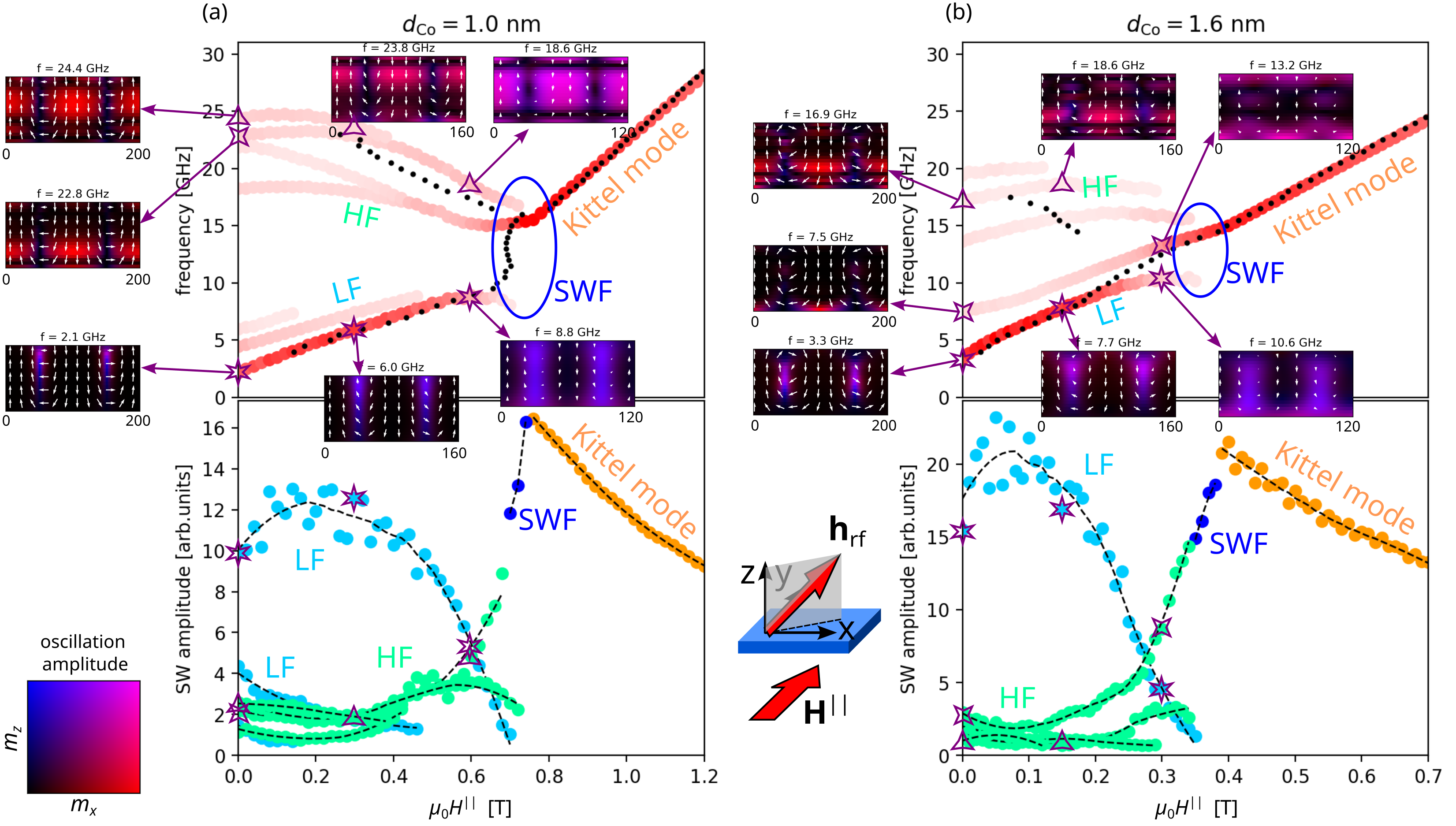}
\caption{Resonance response calculated by micromagnetic simulations, for (a) $\dCo=1.0$~nm, (b) $\dCo=1.6$~nm. For each field value, the resonance frequencies and amplitudes extracted from the simulated spectra are shown in the top and bottom panels, respectively. In the top panels, the red color intensity of each point also encodes the amplitude of the corresponding resonance peaks, whereas in the bottom panels, different colors distinguish the different mode branches:
Kittel mode, high-frequency (HF), low-frequency (LF), and neighborhood of spin wave freezing (SWF). The black dots in the upper plots are the experimental VNA points, from Fig.~\ref{figthree}. The black dashed lines in the bottom plots are guides for the eye. The insets illustrate the amplitude of the localized magnetization oscillations within the domain structure, indicated by the various open symbols at the corresponding magnetic field values, oscillation frequencies, and amplitudes. The white arrows in the insets represent the static magnetization distribution (in the $xz$ plane only, identical to the ones in Fig.~\ref{figfive}). The mixed red and blue color components of the background correspond to the amplitudes of $m_x$ and $m_z$, respectively, according to the legend in the bottom-left corner of the figure. Inset in the central part presents the field configuration.}
\label{figsix}
\end{figure*}

The micromagnetic simulations of the FMR were performed by exciting the system with a spatially uniform (corresponding to $k = 0$) microwave field with a temporal \textit{sinc} pulse, which triggers all frequencies in a given range simultaneously. The ac field was oriented along the (111) direction to excite all magnetization components equally (see the inset in the central part of Fig.~\ref{figsix}). The VNA-FMR spectrum was calculated by the fast Fourier transform (FFT) of the temporal dependence of the average magnetization. As the dc field was applied along the $y$ direction, the oscillations of the $\mx$ and $\mz$ components were taken into account. The calculated $f(H)$ dependences, shown in the top panels of Fig.~\ref{figsix}, reproduce the experimental VNA-FMR results well. The color intensity visualizes the resonance mode amplitude (a sum of squared $\mx$ and $\mz$ amplitudes), with only points above a threshold displayed. This mimicks the experimental sensitivity by showing only the most efficiently excited modes. Crossing $\Hspar$ marks a qualitative transition: the uniform Kittel mode gives way to multiple branches, including a distinct low-frequency branch and multiple higher-frequency branches. The amplitudes are presented in the bottom panels of Fig.~\ref{figsix}, in similar style as in Fig.~\ref{figthree}. Above $\Hspar$, the amplitude increases as the field decreases toward the critical value, in agreement with the experimental data, and expectations for the SW freezing effect. Below $\Hspar$, the amplitude in the low-frequency branch gradually increases with decreasing field, reaching a maximum at moderate fields  ($\sim$0.3~T for $\dCo = 1.0$~nm, $\sim$0.1~T for $\dCo = 1.6$~nm). As the phase transition is approached, the SW spectrum progressively softens and the system becomes increasingly susceptible to microwave excitation. The effect is most pronounced for the softened mode at the critical wave vector $k = \kc$ (see below for more details), which ultimately drives the instability, but the dynamic susceptibility grows more broadly across $k$, so an enhanced response can also appear in the uniform ($k = 0$) signal measured by VNA-FMR \cite{Ki23}.

To identify the resonant modes below $\Hspar$ and clarify the origin of the intensity trends, we analyze the spatial mode profiles by computing FFT of the time-dependent magnetization for each simulation cell. The results are shown in Fig.~\ref{figsix} as spatially $(x,z)$-dependent maps of the oscillation amplitude of the dynamic $m_x$ and $m_z$ components, corresponding to specific magnetic field values and frequencies within the branches. Both components are visualized using additive mixing of reddish and bluish colors. Oscillations of the component parallel to the $z$-axis are largely suppressed in regions where the static magnetization is aligned with $z$ and $m_z$ is close to saturation. 
An analogous restriction is valid for the $m_x$ component oscillations, which are not possible where the magnetization is saturated along $x$.

The mode from the lower-frequency branch (LF) exhibits amplitude localized predominantly within the domain walls. This localization reflects the nature of the mode as an oscillation of the domain wall itself -- a picture that becomes clearer when examining the dispersion relations discussed in the subsequent paragraphs. 

The higher-frequency modes (HF) show more complex profiles, with amplitude distributed mainly in the domains. Particular HF modes are concentrated at either the multilayer surfaces or the central region of the domains. The relative contributions of $m_x$ and $m_z$ oscillations differ between the branches and also vary with the magnetic field. 
Within the domains, at zero field the equilibrium magnetization is oriented along the $\pm z$ direction. Therefore, in the present representation, which shows the dynamic $m_x$ and $m_z$ components, only the $m_x$ oscillation amplitude is visible. As the field increases, the magnetization gradually tilts toward the $y$ direction, making the $m_z$ oscillation component possible and increasingly pronounced.
In LF mode, within the N\'eel wall, where the magnetization is oriented along the $x$ direction, only oscillations of $m_z$ are observed, whereas within the Bloch part of the wall (where the magnetization is oriented along the $y$ direction), oscillations of both $m_x$ and $m_z$ are possible. While increasing the field, the region of the domain wall becomes wider, and a larger volume of the sample contributes to the LF mode. Simultaneously, the amplitude of local oscillations in the cells decreases, and these two opposite factors make the field-dependence of 
effective oscillation amplitude (i.e.~oscillations of the averaged magnetization) nonmonotonic, as presented in the bottom panels of Fig.~\ref{figsix}.

To directly observe SW freezing, one needs access to the critical wave vector $\kc$ where the dispersion minimum reaches $f = 0$. Theory \cite{Ki23} and analysis of the instability regimes \cite{Sobucki25} predict $\kc\approx 50-60$~rad/$\mu$m for our samples, which lies beyond the BLS range (up to $\sim$20~rad/µm). Moreover, the FMR spectra in the domain state---with multiple branches emerging below $\Hspar$---cannot be fully understood without examining the underlying band structure: the periodicity of the stripe domains leads to Brillouin zone folding, mode splitting, and band gaps that directly determine the observed resonance frequencies. We therefore employ micromagnetic simulations to reveal the full dispersion relation, validating the results against BLS measurements in the accessible $k$-range. This approach allows us to connect the experimentally observable quantities---saturation field, domain period, FMR spectra, and partial dispersion---with the freezing process occurring at wave vectors inaccessible in direct measurement.

\begin{figure*}[tbh!]  
\centering
 \includegraphics[width=0.98\textwidth]{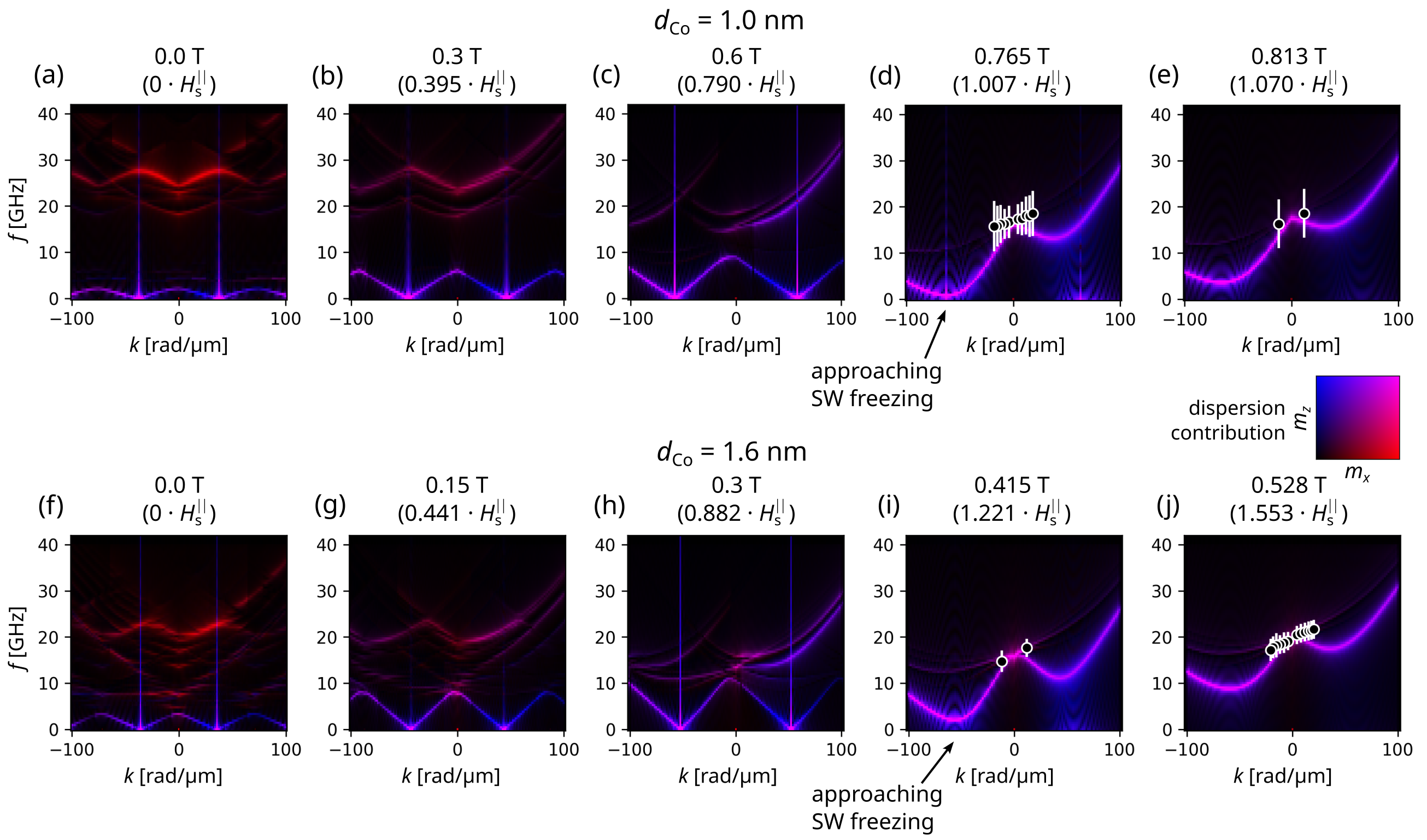}
\caption{SW dispersion relations, calculated by micromagnetic simulations, for (a--e) $\dCo=1.0$~nm, (f--j) $\dCo=1.6$~nm. In (a--c) and (f--h) the field values are below the saturation, and the complex dispersions of domain structure oscillations are observed. In (d, e) and (i, j) the sample is saturated, however in (d) and (i) the field is slightly above $\Hspar$, and the sample is close to SW freezing. 
The mixed red and blue color components correspond to the dispersions of $m_x$ and $m_z$ oscillations, respectively, according to the legend at the right edge of the figure. The experimental values of $(k,f)$ from BLS are plotted with white open symbols; their vertical whiskers correspond to the FWHM of the registered BLS peaks.}
\label{figseven}
\end{figure*}

To access the full dispersion relation for SWs propagating along the $x$-axis (across the domain structure), each simulation was first initialized by relaxing the magnetization into the equilibrium configuration corresponding to the minimum-energy state of the system at the given field. We then triggered magnetization dynamics by a microwave field pulse of both temporal and spatial \textit{sinc} shape ($h_\mathrm{rf} \propto \mathrm{sinc}(2\pi f_\mathrm{cut} t) \mathrm{sinc}(k_\mathrm{cut} x)$, with $f_\mathrm{cut}$ and $k_\mathrm{cut}$ being the cut-off frequency and wave vector), oriented along (111) direction (see the inset in the central part of Fig.~\ref{figsix}), allowing simultaneous excitation of all magnetization components with a broad range of frequencies and wave vectors. The dispersion relation was then calculated as the absolute value of the 2D FFT of the spatio-temporal dependence of the $m_x$ and $m_z$ components of magnetization. The resulting dispersion relations for both samples at various field values are shown in Fig.~\ref{figseven}.

For $\Hpar > \Hspar$, the dispersion relation is relatively simple, with dominant intensity in the lowest band. 
Here, the band carrying most of the spectral weight corresponds to the Damon-Eshbach (DE) mode and, although it may cross thickness-quantized modes, is the lowest-frequency branch for most $k$ and hosts the global minimum of the dispersion.
In addition, weaker higher-order bands are visible, associated with thickness-quantized modes, which appear at similar frequencies near $k = 0$. On both sides of each spectrum ($k < 0$ and $k > 0$) there is a minimum where d$f$/d$k$, and thus the group velocity, vanishes. The asymmetry and different frequency values at these minima are expected to arise from DMI. As the field decreases toward $\Hspar$, the minimum on one side (selected by the DMI sign) approaches $f = 0$, where the phase velocity also vanishes. Simultaneously, the amplitude within this minimum strongly increases, assisted by a distinct increase also for a broad range of $k$, as was observed here both in the experiment and the simulations for $k=0$. 
When this minimum reaches $f = 0$, the system reaches the SW-freezing threshold associated with this mode, marking the critical point at which the stripe-forming instability sets in. The detailed dynamics in this regime is discussed in \cite{Ki23, Sobucki25}.

Below $\Hspar$, the dispersion relations change drastically and become periodic in $k$-space, as expected for a magnonic crystal formed by the stripe-domain pattern. The lowest band of the uniform state splits into two: a Goldstone-like mode with $f = 0$ at $k = \kc$ \cite{Gr22}, and a second band that largely retains the Damon-Eshbach-like character of the former lowest mode. The periodicity introduces a Brillouin-zone structure, so higher-frequency branches (including those connected to thickness-quantized modes) are back-folded into the first zone and can produce additional crossings and mode signatures near $k = 0$. As the field decreases further, the frequency gap between the Goldstone-derived band and the higher-frequency bands grows, while the bandwidth of the Goldstone-derived band narrows significantly. This narrowing originates from the increasing domain period: as domain walls become more widely spaced, their mutual coupling decreases, reducing the dispersion of the wall-localized mode. Since this band is pinned to $f = 0$ at $k = \kc = 2\pi/pc$ and the frequency increases toward $k = 0$, reduced dispersion directly translates to a lower resonance frequency at $k = 0$---consistent with the FMR observations in Fig.~\ref{figsix}. 

Additionally, the branches associated with thickness-quantized modes, which were nearly degenerated for $\Hpar > \Hspar$, progressively split as the field is reduced. This is particularly visible at $k = 0$ and contributes to the increasing number of resonances observed in FMR. Importantly, the back-folded replicas are often much weaker in intensity (depending on field), so some folded branches appear only as faint features in the spectra. Between the Goldstone-derived band and the intense higher-frequency bands, further modes become visible, corresponding to SWs quantized through the thickness of the domain walls.

The calculated dispersion relations can be compared with BLS data. A good agreement is observed, and in both cases a similar DMI-induced asymmetry is visible. The calculated dispersions have narrow, well-resolved bands, because a low damping parameter ($\alpha = 0.01$) was used. For real Co films, an order-of-magnitude higher damping values are expected, and indeed the experimental FWHM values (width of gray shading in Fig.~\ref{figfour} and vertical whiskers in Fig.~\ref{figseven}) extend widely, covering bands that are distinct in simulations. Such bands, being spectrally broad in experiment, overlap and appear as single peaks at some effective central frequencies. Despite this limitation, the agreement in the accessible $k$-range validates the simulations and supports their predictions at higher wave vectors where direct measurements are not possible.

The differences between the two samples can now be understood comprehensively. The critical field $\Hspar$ is determined jointly by $Q$ and $D$: a higher PMA lowers the Kittel-mode frequency as well, but much more strongly deepens the dispersion minimum at finite $k$, while stronger DMI increases the asymmetry, causing one minimum to reach $f = 0$ at higher fields. Consequently, the sample with higher $Q$ and $D$ ($\dCo = 1.0$~nm) exhibits $\mu_0\Hspar \approx 0.76$~T compared to 0.34~T for $\dCo = 1.6$~nm. This has several implications for SW dynamics. The stripe domain state extends over a wider field range for $Q > 1$ (0 -- 0.76~T versus 0 -- 0.34~T). Both samples show a similar total change of the domain period ($\sim$110~nm at $\Hspar$ to $\sim$170--180 nm at $\Hpar$ = 0), but this evolution is compressed into a narrower field range for $Q$ < 1, resulting in faster period change and more rapid evolution of the Brillouin zone with field. The thicker magnetic layer supports more thickness-quantized modes, but these are more closely spaced in frequency and tend to overlap, yielding fewer distinctly resolved FMR bands. Comparing the Goldstone-derived band at $\Hpar = 0$, where domain periods are nearly identical, it remains narrower for $Q > 1$. This reflects stronger localization of the domain-wall mode in the samples with higher PMA: deeper confinement reduces spatial overlap between neighboring walls, suppressing their coupling even at comparable separations. Importantly, when dispersion relations are compared at equivalent reduced fields $\Hpar/\Hspar$ rather than absolute field values, both samples exhibit qualitatively identical behavior -- confirming that the underlying physics is universal, with quantitative differences arising solely from material parameters.

The simulations reveal that at $\Hpar = \Hspar$, the dispersion minimum reaches $f = 0$ with simultaneously vanishing SW group and phase velocities---the defining conditions for SW freezing \cite{Ki23}. While decreasing $\Hpar$, approaching $\Hspar$, the SW amplitude increases (what is consistent with the experimentally observed amplitude increase), and the system undergoes the transition from SWs into a periodic stripe-domain state at $\Hspar$. In the simulations, the emerging critical period $\pc$ just below $\Hspar$ is set by the critical wavev ector, $\pc\approx 2\pi/\kc$, and then increases as the field is reduced further, reaching 170 -- 180~nm at $\Hpar = 0$ in agreement with the remanent-domain periods measured by MFM. This supports the picture that the soft mode at $k=\kc$ seeds the stripe pattern, while the equilibrium period evolves with field within the domain state. 

\section{Summary}

A comprehensive study of [Re/Co($\dCo$)/Pt]$_{20}$  multilayers was performed, covering magnetization statics and dynamics through both experimental and simulation approaches. Modifying the magnetic anisotropy by changing the Co thickness $\dCo$ enabled the creation of nanostructures with $Q>1$ and $Q<1$. The high number of the Co layers repetitions enabled the existence of hybrid domain structures with an out-of-plane magnetization component even for samples with $Q<1$. Asymmetric Co layer interfaces were responsible for the existence of DMI. The nanostructures were investigated in magnetic field applied in the sample plane above saturation, where magnetization dynamics (SWs) were detected. Decreasing the field down to $\Hspar$, the SW amplitude was increasing, which was consistent with approaching SW freezing. As the field is decreased below $\Hspar$, the magnetization excitation spectra change when SWs freeze into domain structures. Pronounced VNA-FMR spectral changes are visible for $\Hpar<\Hspar$: two SW branches and a different inclination in the main branch compared to that of the Kittel mode (existing for $\Hpar>\Hspar$). From micromagnetic simulations, the lower-frequency branch was identified as the oscillations within the domain walls, while for the higher-frequency branch as the oscillations within the domains.
The consistency between independent experimental methods (SQUID, MFM, VNA-FMR, BLS) and simulations---each accessing different aspects of the freezing process---together with the universal scaling behavior observed across samples with different $Q$ and $D$ values, provide comprehensive evidence for SW freezing at the spin reorientation transition.
The similar behavior of both samples, when compared at reduced fields $\Hpar/\Hspar$, indicates that SW freezing near the spin reorientation transition is governed by a universal mechanism, while $Q$ and $D$ primarily set the quantitative scale of the effect. 

Demonstration of SW freezing in material with DMI can have important consequences. DMI-induced nonreciprocity makes the SW dispersion asymmetric, $f(k)\neq f(-k)$, so that waves propagating in opposite directions have different frequencies, phase and group velocities. In contrast to the DMI-free case, where the reciprocal dispersion softens symmetrically and both $\pm k_c$ minima reach $f=0$ together, with DMI only one minimum (selected by the sign of $D$) reaches $f=0$, so the freezing becomes direction-selective. 
Moreover, unlike the DMI-free case studied in Ref.~\cite{Gr22}, where the transition is accompanied by a clearly resolved Higgs-like band emerging from $f = 0$ at $k_c$, we do not resolve a similarly distinct Higgs-like band in the present system. This points to an interesting open question: how DMI-induced nonreciprocity modifies the SW band structure of stripe domain patterns near the spin-reorientation transition, and whether, and if so how, Higgs-like excitations manifest themselves in stripe-domain systems with DMI.
Thus, the presence of DMI not only makes SW freezing direction-selective, but may also modify the collective excitations associated with the transition. Direction-selective freezing opens interesting perspectives, for example for SW amplification and the observation of antimagnons \cite{Sobucki25}.  
A direct experimental confirmation of SW freezing is still demanding e.g.~using the BLS techinque, due to the limited range of available wave vectors in typical BLS setups. 
Application of shorter electromagnetic  wavelength (from the ultraviolet range in the case of the discussed samples) is required for studies with $k$ close to $\kc$.
Another approach is to select samples with properly larger (approximately an order of magnitude) $\pc$, i.e.~shorter $\kc$, inside the wave vector range available in BLS. According to our theory \cite{Ki23}, this should be possible for Co multilayer thickness of the order of 0.5~$\mu$m and $Q\approx 10^{-2}$, alternatively another material, like permalloy or yttrium iron garnet, of properly selected magnetic parameters \cite{lesniewski26}.

\section*{Acknowledgement}

This work was supported by the National Science Centre in Poland under the Project No.~2020/37/B/ST5/02299. 
Micromagnetic simulations were performed at the Computational Centre, University in Bia\l ystok.
The authors thank Dr.~Sukanta Kumar Jena for his contribution in the experimental part of this work.

%

\end{document}